\documentclass{article}
\usepackage{amsmath,amssymb,latexsym,amsfonts,color,enumitem, amsthm}%, authblk,, tabularx, url}
\usepackage[margin = 2.5 cm]{geometry}
\usepackage{graphicx}
\usepackage[T1]{fontenc}
\usepackage{chicago}
\usepackage{url}
\usepackage{cite}
\usepackage{subfigure}
\usepackage{epstopdf}
\DeclareGraphicsExtensions{.jpg, .pdf, .png, .gif, .jpeg}

\newtheorem{theorem}{Theorem}
\newtheorem{corollary}{Corollary}

\newtheorem{lemma}{Lemma}

\newtheorem{proposition}{Proposition}
\newtheorem{remark}{Remark}

\newcommand{\st}{\operatorname{s.t.}}
\newcommand{\E}{\operatorname{E}}
\newcommand{\Var}{\operatorname{Var}}
\newcommand{\Cov}{\operatorname{Cov}}

\newcommand{\R}{\operatorname{R}}
\newcommand{\Diag}{\operatorname{Diag}}

\newcommand\numberthis{\stepcounter{equation}\tag{\theequation}}

\title{Static Hedging of  Weather and Price Risks in Electricity Markets}

\author{Javier Pantoja Robayo\\
 School of Economics and Finance,  Universidad EAFIT. Medellin, Colombia \\
              {\tt jpantoja@eafit.edu.co}
         \and
        Juan~C. Vera\\
 Tilburg School of Economics and Management, Tilburg University, The Netherlands \\
           {\tt j.c.veralizcano@uvt.nl}
}

\begin{document}

\maketitle

\begin{abstract}
We present the closed-form solution to the problem of hedging price and quantity risks for energy retailers (ER), using financial instruments based on electricity price and weather indexes. Our model considers an ER who is intermediary in a regulated electricity market. ERs buy a fixed quantity of electricity at a variable cost and must serve a variable demand at a fixed cost. Thus ERs are subject to both price and quantity risks. To hedge such risks, an ER could construct a portfolio of financial instruments based on price and weather indexes. We construct the closed form solution for the optimal portfolio for the mean-VaR model in the discrete setting. Our model does not make any distributional assumption.

{\bf keywords:} Static Hedging,  Risk Mitigation, Weather Hedging, Energy Markets.

\end{abstract}

\section{Introduction}

The electric power sector includes the generation, transmission, distribution and commercialization of electric power. Regulation in the Whole Sale Power Market (WPM) also created the figure of {\em Load Serving Entities (LSE)} , which are intermediary agents whose purpose is to make competition dynamic and to provide the final customers with different ways to access competitive prices in the electric market of whole sellers. Regulation in the market allows these agents to sell electric power to their customers through contracts which are endorsed by the electric generators to guarantee supply to the users. LSEs can take endless risks and, in the case of bankruptcy, they have no assets to lose.
Electric power generators have warned of the risk that the existence of agents, who had agreed on long-term contracts without a real electric endorsement and used the electric financial market as an instrument to comply with their contractual obligations, could have on the future feasibility of the electric wholesale market. In times of low prices, the LSEs have probably not had any difficulty to comply with their obligations with the electric financial market and signed contracts.  On the contrary, in times of high prices, LSEs would be surely facing financial difficulties due to their losses and would opt out of businesses in detriment of the wholesale market and their customers.

In this paper we consider {\em regulated} LSEs which are denominated {\em Energy retailers (ER)}. An energy retailer (ER) procures power from the wholesale market at the spot price and resells it to consumers exhibiting variable demand. ERs  have to buy electricity at a variable price set by the supply and demand equilibrium and  serve a volatile demand at predetermined fixed prices. Typical ERs meet their obligations through combinations of long-term contracts, wholesale purchases and self-generation. The regulated demand is inelastic, and each ER has the obligation to deliver electricity on demand at the fixed price without failure, independent of costs.

Due to electric power features ERs' profit is exposed to price and volumetric risks. The ER could hedge price risks by using financial instruments based on electricity price which are available in the financial markets. To hedge volumetric risks
storage of electric power is  currently  not an option, and electricity volume-based  financial instruments do no exists. As price and quantity in the electricity markets are correlated with weather, financial instruments based on weather could be used to hedge price and volumetric risks.

In this paper optimal hedging strategies to price and weather risks factors are proposed. The strategy consist of mitigating  ER risk by constructing a contingency claim based on price and weather derivatives. The use of weather derivatives offers the chance to hedge against weather related risks in electric power markets. Companies hedge their portfolios against unexpected weather variations using contracts that are not correlated with classical financial assets, but are based on weather-linked indexes instead. This contracts allow the agents to transfer their risks exposures to financial markets, using financial instruments to hedge price and quantity fluctuations via climatic fluctuations.

We consider the hedging of price and weather risks in a discrete setting where price, quantity and weather index take values on a discrete set. We construct a contingency claim based on price and weather-linked derivatives, without any assumption on the underlying distributions.  We construct such claim, optimizing the Mean-variance utility function for the ER.
Mean-variance \citeN{Mark} is the basis of the modern portfolio allocation strategies, leading the field of Decision Making in Financial Economics.
Due to its simplicity mean-variance is widely used as an approximation to more general expected utility functions (see,  \citeN{morone}). This paper exploits the convexity of the Mean-Variance utility function, which makes possible to find a global optimum to the posed optimization model (\cite{Boyd}).

\subsection{Contribution}

We construct a closed-form solution in discrete setting for the static hedging strategy for  ERs whose net profits are exposed to price and quantity risks associated with weather fluctuations.

In Theorem \ref{thm:sol} we provide an analytical closed form solution in terms of the input data, which is the probability of observing a given price-quantity-weather index combination and the ERs risk aversion coefficient. We also analyze this optimal solution. In Corollary \ref{cor:twoHedge} we show a `two fund theorem' result; namely, for any risk averse  ER  the optimal contingent claims can be written as a combination of two `basic' contingent claims. In Section \ref{sec:phi=psi} we show that when the \emph{risk free} measure and the \emph{real world} measure coincide, the optimal solution is unique, independent of the ER's level of risk aversion; in this case, there exist contingent claims which maximize the expected profit and minimize the risk at the same time. In Section \ref{sec:indCase} we show that when weather and price are independent  the price contingent claim and the weather contingent claim could be constructed by separately finding the optimal price contingent claim  and the optimal weather contingent claim.
Finally in Section \ref{sec:efFront} we characterize the efficient frontier of the model.

In practical terms, the discrete setting is more appropriate than the continuous one.
First, the discrete setting allows a non-parametric approach, where any general discrete distribution could be used, including the case when dependence between price and weather occurs. Notice that this in contrast to the continuous setting, where the existing results depend on the type of distribution underlying the variables of interest (e.g. \cite{Brik_Ron}, \cite{Lee-Oren}) which requires a (semi-) parametric approach.

Second, the set of assets traded in the market is a discrete set, and thus is more natural to assume that the available data is discrete in nature. Third, it is  easier to interpret the outputs of the discrete model. In particular, is much simpler to replicate an optimal contingent claim in the discrete setting (see: \citeN{xiasu}).

Our methodology  allows the use Quadratic Optimization (QO), which is an optimization model that is efficiently solvable. Notice that even though we obtain a closed form solution, the existing efficient solvers for QO offer the possibility for extensions to our model which could include more constrains.

The rest of the paper is organized as follows. In the rest of this Section the preliminaries are presented. In Section \ref{sec:OptHed} our solution to the hedging portfolio problem is presented. In Section \ref{sec:numerical} numerical experiments are performed to illustrate our results, and Section \ref{sec5-2} concludes.

\subsection{Weather Risk and Weather Derivatives}

Earnings obtained by weather-sensitive industries are affected by weather anomalies which is the case of energy industries (\citeN{Dut-2002}). Weather risk in electric power markets is significant; unexpected changes in weather or hydrology could affect quantity and price fluctuations. In countries with seasons, random movements in temperature affect electric power demand. \citeN{Huis} shows that the
difference between actual and expected temperature significantly influences the probability on a spike in day-ahead prices in power markets.
Some tropical countries are also affected by hydrological conditions, demand is correlated with weather fluctuations and the electricity  production might be based on hydro-generation.

Evidence from the Colombian and the US markets shows that climate variation affects spot price and quantity behavior. The wholesale
spot price in Colombia exhibits volatile behaviour, for instance in November 2010 average daily price per KWh was 14.1\% less than in November 2009 due to La Ni\~{n}a phenomenon (138 COPs per KWh in November 2010, and 160.7 COPs per KWh in November 2009).
In the generation side, there was 19.8\% more capacity in 2010 after the effects of 2009 El Ni\~{n}o had diminished (75.2\% in 2010 and 55.4\% in 2009). Similarly, in the US, the spot price in PJM Market reached \$141 per MWh in the
summer of 2010, while the previous summer the  price range was  \$31 - \$80. In the summer of 2010 load in the PJM system was 135
GW, while for the previous summer was around 186 GW.  For more detail see \citeN{Pant11}.

Weather derivatives could be used to hedge unexpected changes in weather. These kind of derivatives were first launched in 1996 in the Chicago Mercantile Exchange (CME), in the United States, as a mechanism of protection against weather anomalies; since then, they are used by companies to transfer climate-related risk to capital markets. The underlying of weather derivatives are the weather fluctuations, and thus weather derivatives are use to smooth out weather related revenue fluctuations for companies, transforming non-tradable risk into tradable financial securities, which brings financial and commercial benefits (\citeN{Jew-2004}). Weather derivatives are based on indexes of temperature, such as the Cooling-Degree-Days (CDD) and the Heating-degree-Days (HDD) at the CME.

The weather derivatives market does not have yet an effective pricing model. Constructing such model is difficult, due to the incompleteness of the market and the fact that
weather derivatives are basically a speculative security as weather indexes are not a tradable commodity or a delivery asset.
\citeN{Rich-2004} presented a pricing model based on a temperature process constructed from a mean-reverting Brownian motion. \citeN{Chaum} considered a model where under an equilibrium condition, the market price of risk is uniquely determined by a backward stochastic differential equation;  this stochastic equation is translated into a semi-linear partial differential equation which is solved using two simple models for sea surface temperature.

\subsection{Risk hedging in the Electricity market}
The hedging of price, quantity and weather risks in the Electricity market have been considered in recent literature.
Several authors follow Markowitz's \emph{mean-var} methodology \citeN{Mark} to construct hedging strategies.
\citeN{Woo-2004}, studied the interaction between stochastic consumption volumes and electricity prices, and proposed a mean-var model to determine optimal hedging strategies.
\citeN{NAK} develop a hedging strategy for risk mitigation using forward contracts based on price, their formulation is an extension of the formulation given on \citeN{mck} to obtain the optimal hedge ratio considering price and quantity risks. \citeN{Oum-Oren} extend the model of \citeN{NAK} to include price derivatives in the hedging strategy.
\citeN{Lee-Oren} extend \citeN{Oum-Oren} model, via the inclusion of forwards based on weather-linked indexes. \citeN{Pant11} and \citeN{Brik_Ron} consider the case of risk hedging assuming independence between price and weather index.

Other Risk measures have also been considered. For instance,
\citeN{Veh-Keppo} suggests using \emph{value at risk} as risk measure, when managing risk in the electricity market.
Hedging of ER's  risk under value at risk using price derivatives has been considered by \citeN{Oum-Oren}, \citeN{Kle-Li}, \citeN{Woo-2004}, and \citeN{Wagner-2003}.

\subsection{The Hedging model} \label{sec.SolMeth} \label{sec:FixPoint}
We study a one stage model for the hedging problem for one ER.  The ER buys electricity at the spot price $p$ and sells it at fixed retail price $r$ at a fixed time $T$. Therefore the ER's profit, at time $T$, from serving the customers' demand $q$ is $y(p,q)=(r-p)q$.
 The retail price $r$ is fixed and known, while the demand $q$  and the spot price $p$ are random.

To hedge risks due to price and quantity fluctuations, the ER constructs contingent claims $x_p(p)$ and $x_w(w)$, which represent the payoffs of financial portfolios based on electricity price and weather-index instruments respectively.
These contingent claims are functions of the electric spot price and the weather index respectively. Notice that we work directly with contingent claims, and disregard the details on how to construct replicating portfolios composed by vanilla and derivatives.  For such details, in discrete setting, see \citeN{Leung-loring}.

To find the optimal contingent claims we use the  following optimization model proposed by  \citeN{Pant11}  (see also, \cite{Brik_Ron}).
\begin{eqnarray}\label{eq:main}
\max_{x_p,x_w}& \E^{\psi}[y(p,q)+x_p(p)+x_w(w)] - a \Var^\psi[y(p,q)+x_p(p)+x_w(w)]\\
\nonumber \st & \E^{\phi}[x_p(p)]=0 \\
\nonumber  & \E^{\phi}[ x_w(w)]=0.
\end{eqnarray}
In model~\eqref{eq:main} the probability measure $\psi$ supported on $\{(p,q):p \ge 0,\,q \ge 0\}$ represents (the ER beliefs on) the real distribution of the realization of $p$ and $q$ at time $T$, and $\phi$ is a risk neutral probability measure. Notice that we do not assume $\phi$ to be unique, since the electric power market is incomplete. The expectations $E^{\psi}[.]$ and $E^{\phi}[.]$ denote expectations under the probability measure \textbf{$\psi$} and \textbf{$\phi$}, respectively.

The objective of~\eqref{eq:main} is to maximize the Mean-Variance utility function $U(Y) = Y - a(Y-E^\psi[Y])^2$
 of the ER's hedged profit at time $T$. Model~\eqref{eq:main} seeks to maximize the ER's expected utility, under the constrain that the cost of constructing the financial portfolios with contingent claims $x_p(p)$ and $x_w(w)$ is zero, that is those portfolios are \emph{zero-cost}.

Notice that in this context having two separate zero-cost portfolios with contingent claims $x_p(p)$ and $x_w(w)$ is equivalent to having one zero-cost portfolio with contingent claim $x_p(p) + x_w(w)$. We prefer to keep them separated in the constrains of model \eqref{eq:main}, as this simplifies our exposition.

\citeN{Oum-Oren}, develop a static hedging estrategy based on composition of portfolio price-based financial energy instruments, where only a contingent claim $x_p(p)$ is considered. \citeN{Lee-Oren} extend \cite{Oum-Oren} model to one that includes forwards on weather index, which is equivalent to model \eqref{eq:main} restricting the weather contingent claim to be of the form $x_w(w) = \alpha w$, where $\alpha$ is an optimization variable.
Considering a contingent claim $x_w(w)$ of general form offers the possibility of a better hedge, as shown in our numerical examples (Section \ref{sec:numerical}). Also the consideration of general contingent claim makes the optimization problem more complex. \citeN{Pant11} and \citeN{Brik_Ron}  derive first order conditions for model \eqref{eq:main} and use them to find a closed form solution in the case when the price is not correlated to the weather index.

All the previous works assume the distributions $\psi$ and $\phi$ are continuous and known.
In contrast, we work on a discrete setting. That is we assume the distributions $\psi$ and $\phi$ to be discrete and known.
We present a closed form solution, derived using QO.
The existence of efficient solvers for QO opens the possibility to modelling extensions. For instance, transaction costs and/or robustness  could be added as linear or second order constraints, which produce models that are still efficiently solvable.
Moreover, the combination of the discrete setting and the QO-based methodology, allows to consider any general discrete distribution, including the case when  price and weather are not independent. Notice that working with continuous distribution requires working on a semi-parametric setting (see e.g. \citeN{Brik_Ron}) while discrete distributions allow to use a non-parametric data-based setting (see Section \ref{sec:numerical}).

\section{Optimal Hedging via Quadratic Optimization}\label{sec:OptHed}

In this section we present a Quadratic Optimization based approach to solve model \eqref{eq:main}.
We assume a discrete setting with $n$ possible prices, $m$ possible values for the weather index, and $\ell$ possible values for quantity. That is, we are given a set $P =\{p_i:i=1,\dots,n\}$ a set $W = \{w_j:j=1,\dots,m\}$ and a set $Q = \{q_1,\dots,q_{\ell}\}$ of possible prices, weather index values and quantities respectively.  In  this setting, the ER- beliefs distribution  $\psi$ is a probability measure supported on $P \times Q \times W$
and the risk-free distribution  $\phi$ is a probability distribution supported on $P \times W$.

Let $\mu_y = \E^{\psi}[y(p,q)]$ be the expected value of $y(p,q)$ under $\psi$, and $\sigma^2_y = \Var^\psi[y(p,q)]$ be its variance. Let $\psi_p$ be the marginal of $\psi$ on $P$, that is $\psi_p(p) = \sum_{q,w}\psi(p,q,w)$. We will abuse the notation and also use $\psi_p \in \R^n$ to denote the corresponding vector of marginal probabilities. Similarly define $\psi_w$. Also, let $\psi_{pw}(p,w) = \sum_{q}\psi(p,q,w)$ be the 2-marginal on $P \times W$ of $\psi$. Again, we will abuse the notation and use $\psi_{pw}$ to denote the corresponding matrix of marginal probabilities.
 Also, for any $p$, define $\mu^p_y(p) = \E^\psi[y(p,q)|p]$, and use $\mu_y^p$ to denote the corresponding vector in $R^n$. Similarly define $\mu^w_y$.

Notice that our decision variables are vectors $x_P$ and $x_W$ indexed by $P$ and $W$ respectively. This vectors represent the contingent claim function value at each $p \in P$ (resp. $w \in W$). More exactly, the price-contingent claim function is $x_p = x_P^T \delta_P$, where $\delta_P$ is a random vector drawn from $\{u \in \{0,1\}^P: \Sigma_p u_p = 1\}$ according to $\psi_p$. I.e. $\delta_P$ is the random 0-1 vector with exactly one 1 in the realized $p$. Analogously $\delta_W$ is defined to obtain $x_w = x_W^T \delta_W$. We obtain then the following straightforward proposition.

\begin{proposition} \label{prop:deltas} Let $\delta_P$ and $\delta_W$ be defined as above. We have then,
\begin{enumerate}
\item $\E^{\psi}[\delta_P]  = \psi_p$ and $\E^{\psi}[\delta_W]  = \psi_w$.
\item $\E^{\psi}[\delta_P \delta_P^T]  = \Diag(\psi_p)$ and $\E^{\psi}[\delta_W\delta_W^T]  = \Diag(\psi_w)$.
\item $\E^{\psi}[\delta_P \delta_W^T]  = \psi_{pw}$.
\item $\E^{\psi}[\delta_P y(p,q)]  = \psi_p \circ \mu_y^p$ and $\E^{\psi}[\delta_Wy(p,q)]  = \psi_w \circ \mu_y^w$, where $\circ$ is the component-wise product of vectors.
\end{enumerate}
\end{proposition}

From Proposition \ref{prop:deltas} we obtain
\begin{align*}
\E^{\psi}[y(p,q)+x_p(p)+x_w(w)]&=
\E^{\psi}[y(p,q)] + \E^{\psi}[x_P^T \delta_P] + \E^{\psi}[x_W^T \delta_W]
= \mu_y +  \psi_P^T x_P + \psi_W^Tx_W.
\end{align*}

And, letting $e_P$ (respectively $e_W$) denote the all-ones vector indexed by $P$ (resp. $W$),
\begin{align*}
\Var^{\psi}&[y(p,q)+x_p(p)+x_w(w)]  \\
&=\Var^\psi[y(p,q)]+\Var^{\psi_p}[x_P^T \delta_P]+\Var^{\psi_w}[x_W^T \delta_W]\\
&\qquad + 2\Cov^{\psi}[x_P^T \delta_P,y(p,q)]+ 2\Cov^{\psi}[x_W^T \delta_W,y(p,q)]+2\Cov^{\psi}[x_P^T \delta_P,x_W^T \delta_W] \\
&= \sigma^2_y+ x_p^T(\E^{\psi}[\delta_P \delta_P^T] - \E^{\psi}[\delta_P]\E^{\psi}[\delta_P]^T)x_p  + x_w^T(\E^{\psi}[\delta_W \delta_W^T] - \E^{\psi}[\delta_W]\E^{\psi}[\delta_W]^T)x_w\\
&\qquad + 2( E^{\psi}[y(p,q)\delta_P] - E^{\psi}[y(p,q)]E^{\psi}[\delta_P])^T x_P + 2( E^{\psi}[y(p,q)\delta_W] - E^{\psi}[y(p,q)]E^{\psi}[\delta_W])^T x_W\\
&\qquad\qquad + 2x_P^T ( E^{\psi}[\delta_P \delta_W^T] - E^{\psi}[\delta_P]E^{\psi}[\delta_W]^T)x_W \\
&= \sigma^2_y+ x_p^T(\Diag(\psi_p) - \psi_p\psi_p^T)x_p  + x_w^T(\Diag(\psi_w) - \psi_w\psi_w^T)x_w + 2x_P^T ( \psi_{pw} - \psi_p\psi_w^T)x_W\\
&\qquad + 2 (\mu_y^p\circ \psi_p -  \mu_y \psi_p )^T  x_P + 2 (\mu_y^w \circ \psi_w -  \mu_y\psi_w )^T x_W.
\end{align*}

We obtain then,
\begin{align*}
&\E^{\psi}[y(p,q)+x_p(p)+x_w(w)] - a\Var^{\psi}[y(p,q)+x_p(p)+x_w(w)] \\
&\qquad=\mu_y - a\sigma^2_y + (\psi_p + 2a c_p)^T x_P + (\psi_w + 2a c_w)^T x_W - a\begin{bmatrix}x_P \\ x_W\end{bmatrix}^T \begin{bmatrix}M_{pp} & M_{pw}\\ M_{pw}^T & M_{ww}\end{bmatrix} \begin{bmatrix}x_P \\ x_W\end{bmatrix}
\end{align*}
where
\begin{align*}
c_p &= \mu_y\psi_p - \mu^p_y \circ \psi_p&
c_w &= \mu_y\psi_w - \mu^w_y\circ \psi_w \\
M_{pp} &= \Diag(\psi_p) - \psi_p\psi_p^T&
M_{ww} &= \Diag(\psi_w) - \psi_w\psi_w^T&
M_{pw} &= \psi_{pw} - \psi_p\psi_w^T. \numberthis \label{def:M}
\end{align*}
Similarly,
\[
 \E^{\phi}[x_p(p)] = \phi_p^T x_P \qquad\text{ and }\qquad
 \E^{\phi}[x_w(w)] = \phi_w^T x_W
\]
where
\begin{equation}\label{def:b}
\phi_p \in \R^n \text{ and }\phi_w\in \R^m
\end{equation}
 are the marginals of $\phi$.

Model \eqref{eq:main} is then equivalent  to
\begin{eqnarray}\label{eqmv1.2}
\mu_y - a\sigma^2_y +  &\displaystyle\max_{x_P,x_W}& \begin{bmatrix}\psi_p + 2a c_p \\ \psi_w + 2a c_w\end{bmatrix}^T\begin{bmatrix}x_P\\x_W\end{bmatrix} - a\begin{bmatrix}x_P \\ x_W\end{bmatrix}^T \begin{bmatrix}M_{pp} & M_{pw}\\ M_{pw}^T & M_{ww}\end{bmatrix} \begin{bmatrix}x_P\\x_W\end{bmatrix} \\
\nonumber
&\st & \phi_p^T x_P=0 \\
\nonumber  && \phi_w^T x_W =0\\
\nonumber  && x_P \in \R^n,\,x_W \in \R^m
\end{eqnarray}
To simplify the notation, define $M =\begin{bmatrix}M_{pp} & M_{pw}\\ M_{pw}^T & M_{ww}\end{bmatrix}$, $B = \begin{bmatrix}  \phi_p & 0 \\0& \phi_w\end{bmatrix}$,  $b = \begin{bmatrix}  \phi_p \\ \phi_w\end{bmatrix}$, $c = \begin{bmatrix}c_p\\c_w\end{bmatrix}$ and $d = \begin{bmatrix}\psi_p \\\psi_w \end{bmatrix}$.

The matrix
$M$ is positive semidefinite. Thus, for a>0  model \eqref{eqmv1.2} is a (convex) QO Problem. For QO the first order conditions (FOCs) imply optimality \citeN{Keller}. In Section 2.1. next, an optimal solution is obtained from the FOCs.

\subsection{Closed form solution in the discrete setting}

In Theorem \ref{thm:sol} we give a closed-form solution to model \eqref{eqmv1.2}. Obtaining a closed form solution is possible as model \eqref{eqmv1.2} only has equality constrains. Adding restrictions on the contingent claims might be desirable in practical situations. For example one might want to constrain the corresponding portfolio to closely follow a given benchmark (see \cite{CPT18} and references therein). Many constraints used in practice can be expressed as linear constraints on the contingent claims and thus when added to model \eqref{eqmv1.2} the model will be still a QO problem which could be solved efficiently. Commercial packages such as Cplex (\cite{CPLEX}) and Gurobi (\cite{gurobi})
could solve QO Problems very efficiently.

\begin{theorem}\label{thm:sol}  Let $a > 0$.
\begin{enumerate}
\item The pair $(x_P,x_W)$ is an optimal solution to model \eqref{eqmv1.2} if and only if there exist $\lambda_p, \lambda_w \in \R$ such that
\begin{align}\label{FOC}
2aM \begin{bmatrix}x_P\\x_W\end{bmatrix} + B \begin{bmatrix}  \lambda_p \\\lambda_w\end{bmatrix} &  =  2ac + d
\text{ and } B^T \begin{bmatrix}x_P\\x_W\end{bmatrix} =0.
  \end{align}
\item  Let $\hat M$ (resp. $\hat b$, $\hat c$ and $\hat d$) be the matrix obtained from dropping from $M$ the last of the $p$-rows (resp. $p$-entries) and the last of the $w$-rows (resp. $w$-entries).
The unique solution to model \eqref{eqmv1.2} is
\begin{align}\label{eq:sol}
  \begin{bmatrix}x_P\\x_W\end{bmatrix}=  \begin{bmatrix}\hat M \\B^T\end{bmatrix}^{-1}\begin{bmatrix}\hat c + \frac{1}{2a}(\hat d - \hat b)\\ 0\end{bmatrix}\\
\nonumber \lambda_p = 1, \qquad \lambda_w =1
\end{align}
\end{enumerate}
\end{theorem}

\begin{proof} To prove the first part, notice that because the feasible set of model \eqref{eqmv1.2} is a polyhedron, the first order conditions \eqref{FOC} are  necessary conditions for the optimal solutions (see, e.g. \citeN{Eus-2008}). Also, as $M$ is positive semi-definite, model \eqref{eqmv1.2} is convex. Also $(x_p , x_w) = (0,0)$ is a Slater point, and thus \eqref{FOC} is sufficient for optimality (see, e.g. \cite{Boyd}).

To prove the second part,  Let $e_p \in \R^n$ and $e_w \in \R^m$ be the  all ones vectors. Multiplying \eqref{FOC} to the left by $\begin{bmatrix}e_p^T &0 \end{bmatrix}$ we obtain
\[2a e_p^T M_{pp} x_p + \lambda_p e_p^T \phi_p  = e_p^T \psi_p + 2a e_p^Tc_p.\]
But, $e_p^T M_{pp} = 0$, $e_p^T \phi_p =1$, $e_p^T \psi_p =1$ and $e_p^T c_p = 0$. Thus $\lambda_p =1$. Similarly $\lambda_w =1$.
Thus \eqref{FOC} is equivalent to
\begin{align}\label{FOC1}
\begin{bmatrix}M\\B^T\end{bmatrix}  \begin{bmatrix}x_P\\x_W\end{bmatrix} &  = \begin{bmatrix} c + \frac{1}{2a}(d - b)\\ 0\end{bmatrix}.
\end{align}

Let $u_1 = {\tiny \begin{bmatrix}e_p \\0 \end{bmatrix}}$ and $u_2 = {\tiny\begin{bmatrix}0 \\ e_w \end{bmatrix}}$. Then $Mu_i =0$, $cu_i = 0$ and $(d-b)u_i = 0$ for $i=1,2$. Thus, dropping the last of the $p$-rows and the last of the $w$ rows we obtain that \eqref{FOC1} is equivalent to
\begin{align}
\begin{bmatrix}\hat M\\B^T\end{bmatrix}  \begin{bmatrix}x_P\\x_W\end{bmatrix} &  = \begin{bmatrix} \hat c + \frac{1}{2a}(\hat d - \hat b) \\ 0\end{bmatrix},
\end{align}
from  where \eqref{eq:sol} follows.
  \end{proof}

\begin{remark}
  Part 2 of Theorem \ref{thm:sol} implicitly assumes that  $\begin{bmatrix}\hat M\\B^T\end{bmatrix}$ is invertible. Generically, this is true, as $\hat M$ is full row rank (see Lemma \ref{lem:rank} below), and the rows of $B^T$ are generically linearly  independent from the rows of $M$, given the nature of the  distributions $\phi$ and $\psi$.
\end{remark}
Computing $\begin{bmatrix}\hat M \\B^T\end{bmatrix}^{-1}\begin{bmatrix}\hat c + \frac{1}{2a}(\hat d - \hat b)\\ 0\end{bmatrix}$ is equivalent to solving the system
$\begin{bmatrix}\hat M\\B^T\end{bmatrix}  \begin{bmatrix}x_P\\x_W\end{bmatrix}   = \begin{bmatrix} \hat c + \frac{1}{2a}(\hat d - \hat b) \\ 0\end{bmatrix}$ which can be solved using $O((n+m)^3)$ operations. In general this systems are solved using Gaussian elimination or QR decomposition. In the case of ill-posed systems robust versions of this methods can be used, which generate good approximate solutions. For more information see, e.g., \cite{Meyer}

For any $a$ let $x^a  = \begin{bmatrix}x_P^a\\x_W^a\end{bmatrix}$ where $x_P^a$, $x_W^a$ are the optimal solutions to model \eqref{eqmv1.2}.
Let
\[
 x^o = \begin{bmatrix}x_P^o\\x_W^o\end{bmatrix} =  \begin{bmatrix}\hat M \\B^T\end{bmatrix}^{-1}\begin{bmatrix}\hat d - \hat b\\0\end{bmatrix} \text{ and } x^\infty = \begin{bmatrix}x_P^\infty\\x_W^\infty\end{bmatrix} =  \begin{bmatrix}\hat M \\B^T\end{bmatrix}^{-1}\begin{bmatrix}\hat c \\ 0\end{bmatrix}
\]
 In Corollary \ref{cor:twoHedge} we show that for any $a>0$ the optimal contingent claim $x^a$ is a linear combination of $x^o$ and $x^\infty$. Corollary \ref{cor:twoHedge} is a sort of Two Fund Theorem (see \citeN{CPT18}). In particular this implies that the optimal hedging portfolio is a combination of the hedging portfolios corresponding to $x^o$ and $x^\infty$.  Notice that $x^\infty$ is the contingent claim minimizing the total profit risk, while $x^o$ is in correspondence to the contingent claim maximizing the expected value of the hedged profit. That is any optimal portfolio can be obtained by combining a risk-efficient  portfolio with a profit maximizing portfolio. Theorems of this type are important as they allow to compute optimal portfolios even when the risk aversion parameter $a$ is not exactly known.

\begin{corollary}\label{cor:twoHedge}
For any $a>0$
\begin{equation}\label{eq:sola}
x^a =x^\infty + \tfrac 1{2a} x^o.
\end{equation}
\end{corollary}
\begin{proof} Follows from the definition of $x^o$ and $x^\infty$ and \eqref{eq:sol}.
 \end{proof}

To finish we present a technical result (Lemma 1) which guarantees the existence of the closed form solution (see Remark 1, and the subsequent paragraph).
\begin{lemma}\label{lem:rank}
The matrix $M$ is psd, with row rank $n+m-2$. In particular, $\hat M$ the matrix obtained dropping from $M$ the last of the $p$-rows and the last of the $w$-rows is full row-rank.
\end{lemma}

\begin{proof}
$M$ is the Variance-Covariance matrix of the vector  $\begin{bmatrix}\delta_p\\ \delta_w\end{bmatrix}$, thus it is psd.
We claim now that all eigenvalues of $M$ are positive except for the smallest two, corresponding to the eigenvectors $\begin{bmatrix}e_p\\0\end{bmatrix}$ and $\begin{bmatrix}e_w\\0\end{bmatrix}$, where $e_p \in \R^n$ and $e_w \in \R^m$ are the respective all ones vectors. To prove this let  $\begin{bmatrix}u\\ v\end{bmatrix}$ be an eigenvector with $0$ as corresponding eigenvalue. We have then
\[VaR[u^T\delta_p + v^T\delta_w] = \begin{bmatrix}u\\ v\end{bmatrix} M \begin{bmatrix}u\\ v\end{bmatrix} = 0,\]
which implies, $u = \alpha e_p$ and $v = \beta e_w$, where $\alpha$ and $\beta$ are scalars.

As $M$ is a psd matrix with exactly two zero eigenvalues, $M$ has row-rank $m+n-2$. Also, from symmetry of $M$,
$\begin{bmatrix}e_p^T & 0\end{bmatrix}$ is a left eigenvector of $M$ with eigenvalue 0, that is $\begin{bmatrix}e_p^T &0\end{bmatrix}M = 0$. This means that the last of the $p$-rows of $M$ is a linear combination of the other $p$-rows. Similarly the last $w$-row is a linear combination of the other $w$-rows. Thus dropping this two rows we obtain a matrix of the same row-rank as $M$, that is $n+m-2$ that has exactly $n+m-2$ rows. I.e. it is full row rank.
 \end{proof}

\subsection{Analyzing the solution when $\phi = \psi$}\label{sec:phi=psi}

Next, we analyze the rather unrealistic case when $\phi = \psi$.  Special attention has been paid to this case in \cite{Oum-Oren}, \citeN{Pant11} and \citeN{Brik_Ron}. \cite{Oum-Oren} noted
 that assuming $\phi = \psi$, when constructing  the price only hedging in the continuous setting,  the optimal contingent claim is $x_p = \mu_y - \mu_y^p$ independent of the value of $a$. Our next result is that, under the assumption $\phi = \psi$ the optimal contingent claims in the general case are also independent of $a$.

\begin{corollary}\label{cor:phi=psi}
Assume $\phi = \psi$. For any $a>0$ the optimal solution to \eqref{eqmv1.2} is independent of $a$.
\end{corollary}
\begin{proof} If $\phi=\psi$, we have that $b=d$ and thus $x^o=0$. Thus, from Corollary \ref{cor:twoHedge} we have $x^a = x^\infty$.
 \end{proof}

As can be seen from the proof of Corollary \ref{cor:phi=psi}, when $\phi = \psi$ the optimal solution to model \eqref{eqmv1.2} is the contingent claim that minimizes the total  risk, for any $a>0$. A limiting argument shows that this solution is optimal also when $a =0$, that is, this contingent claim also maximizes the expected total profit.

\subsection{Solution to the independent case}\label{sec:indCase}

If the price and the weather index are independent, solution \eqref{eq:sol} simplifies further as shown in Corollary \ref{cor:ind}.
\begin{corollary} \label{cor:ind} Assume the price and the weather index are independent according to $\psi$. Then the solution to model \eqref{eqmv1.2} is given by
\begin{align*}
\nonumber  x_P &= \begin{bmatrix}\hat M_{pp} \\ \phi_p^T\end{bmatrix}^{-1} \begin{bmatrix}\hat c_p + \frac{1}{2a}(\hat \psi_p- \hat \phi_p) \\0 \end{bmatrix}   \\
\label{indep} x_W &= \begin{bmatrix}\hat M_{ww} \\ \phi_w^T\end{bmatrix}^{-1}\begin{bmatrix}\hat c_w + \frac 1{2a}(\hat \psi_w - \hat \phi_w) \\0\end{bmatrix}
\end{align*}
\end{corollary}
\begin{proof} If $p$ and $w$ are $\psi$-independent we have $M_{pw} = 0$, and thus using  part 2 of Theorem \ref{thm:sol} the statement follows.
 \end{proof}

\begin{remark}
  Notice that in the case of $\phi$-independence between $p$ and $w$ we have  $\mu^p_y(p) = \E^\psi[y(p,q)|p] = \E^\psi[(r-p)q|p] = (r-p)\E^\psi[q|p] $. Similarly $\mu^w_y(w)  = \E^\psi[y(p,q)|w] = \E^\psi[(r-p)q|w] = (r-\E^\psi[p])\E^\psi[q|w] $.
\end{remark}
In the case of $\psi$-independence between $p$ and $w$, the optimal solution corresponds to solving independently the hedging problem for $p$ and $w$. One can check that in this case, the optimal contingent claim pair $(x_p,x_w)$ given by Corollary \ref{cor:ind} solves the problems of constructing the best contingent claim using information on price only, and information on weather only, separately. This can also be shown by analyzing the optimization model \eqref{eqmv1.2}, if $M_{PW} = 0$ the model separates, that is  model \eqref{eqmv1.2} is equivalent to
\begin{align*}
c_y + \max_{x_P}\ & c_p^T x_P - x_P^T M_{pp} x_P\qquad + &   \max_{x_P}\ & c_w^T x_W - x_W ^T M_{ww} x_W&\quad& \\
\st\ & b_p^T x_P=0 & \st\ & b_w^T x_W =0\\
 & x_P \in \R^n &&x_W \in \R^m.
\end{align*}

Corollary~\ref{cor:ind+=} describes the optimal contingent claims for the independent case when $\phi = \psi$ is assumed. \begin{corollary}\label{cor:ind+=} Assume $\phi = \psi$ and $w$ independent of $p$. Then the optimal contingent claims are  $x_p = \mu_y - \mu_y^p$ and $x_w = \mu_y - \mu_y^w$.
\end{corollary}
\begin{proof}
 This is easily checked directly by plugin in the given $x_p$, $x_w$ and $\lambda_p = \lambda_w = 1$ in part 1 of Theorem \ref{thm:sol}.
 \end{proof}

\subsection{Efficient Frontier}\label{sec:efFront}

We apply now \cite{Mark} concept of efficient frontier to our model for hedging the electric power market. In Section \ref{sec:numerical}, we use the results of this section to compare different solutions to model \eqref{eqmv1.2}. There, we compare the efficient frontier of the solution obtained by wrongly assuming independence between price and weather with the solution obtained by assuming the right level of correlation. By comparing the efficient frontier of this two solutions, we observe that the solution to the general case dominates, in the Pareto sense, the independence-based  one.

\begin{proposition} \label{prop:frontier}For any $a>0$, let $x^a_p$ and $x^a_w$ be the optimal solution for model~\eqref{eq:main}. Then
\begin{align*}
\E^{\psi}[y(p,q)+x^a_p(p)+x^a_w(w)] & = \mu_y  + d^T x^\infty  + \tfrac 1{2a} d^T x^o\\
\Var^{\psi}[y(p,q)+x^a_p(p)+x^a_w(w)] &= \sigma^2_y  - c^T x^\infty + \tfrac{1}{2a}(c^T x^o + (d - b)^T x^\infty) + \tfrac{1}{4a^2}(d - b)^Tx^o.
\end{align*}
\end{proposition}
\begin{proof}
From Corollary \ref{cor:twoHedge} we have $x^a = x^\infty + \tfrac 1{2a}x^o$. Then
\begin{align*}
\E^{\psi}[y(p,q)+x^a_p(p)+x^a_w(w)] %& = \mu_y  +  \psi_p^T x^a_P  +  \psi_w^T x^a_W\\
& =  \mu_y  + d^Tx^a =  \mu_y  + d^T x^\infty  + \tfrac 1{2a} d^T x^o.%&\text{(from \eqref{eq:sola})}\\
\end{align*}
Also, using \eqref{FOC1},
\begin{align*}
\Var^{\psi}[y(p,q)+x^a_p(p)+x^a_w(w)] \hspace{-8em}& \hspace{8em}= \sigma^2_y  - 2  c^T x^a + (x^a)^T M x^a\\
& = \sigma^2_y  - 2c^T x^a + (c + \tfrac{1}{2a}(d - b))^Tx^{a} \\
& = \sigma^2_y  - c^T x^a   + \tfrac{1}{2a}(d - b)^Tx^a \\
& = \sigma^2_y  - c^T x^\infty + \tfrac{1}{2a}(c^T x^o + (d - b)^T x^\infty) + \tfrac{1}{4a^2}(d - b)^Tx^o. \qquad\qquad
\end{align*}
\end{proof}

\section{Numerical Results}\label{sec:numerical}

In this section we evaluate our proposed solution. We use data similar to the one used in \citeN{Lee-Oren}. Namely,  we construct hedging strategies for an ER that charges a flat retail rate of  \$120 per MWh. We assume lognormal distributions of price and quantity.
Under  $\psi$ the three variables, price, weather index and quantity  are distributed according to a multivariate normal distribution in
log price, log quantity, and the weather-index. Under $\phi$ price and weather index are distributed according to a bivariate normal distribution in log price and weather-index. In all cases, except when looking at the efficient frontier,
the risk aversion parameter is fixed to $a = 1.0$.

Our method is set in a discrete frame. So to apply the method, we should have as input discrete distributions. To do this,
the given continuous distributions are discretized using a grid with $n=100$ points equally spaced from $\mu - 3\sigma$ to $\mu + 3\sigma$ in each of the coordinate axis (log price, log quantity and weather-index), where $\mu$ and $\sigma$ are respectively the mean and the standard deviation of the projection of the corresponding multivariate normal distribution in the given axis. So in total $n^3$ triples  (price, quantity, weather-index), each one with a probability value assigned are obtained as input. We calculate joint distributions of $p$, $q$ and $w$ according with the marginals $\Psi_{p}$ and $\Psi_{w}$.

To produce the output in each case, first contingent claims $x_p^*$ and $x_w^*$ are obtained by solving the corresponding case (using Theorem \ref{thm:sol} or one of its corollaries). The distribution of the hedge profit is then plotted computing the hedged profit for each of the $N = n^3 = 1000$ data points in the input and using the matlab function {\tt ksdensity} which allow us to produce profit distributions plots for each strategy.

\subsection{Independence case}

In the independence case, it is possible to use the analytical solution for the continuous case obtained by \citeN{Pant11} and \citeN{Brik_Ron}. We compute this \emph{continuous solution}, using as input the distributions' parameters given for the independence case.
Then we compare the analytic continuous solution against our \emph{discrete independence solution}. This solution is obtained  using Corollary \ref{cor:ind}.
Following \citeN{Lee-Oren} we use the parameters given in Table~\ref{tab:ind} for the input distributions
\begin{table}[h]
\caption{Data parameters for the independence case.}
\label{tab:ind}\medskip
\begin{tabular}{rlll}
Under $\psi$ : &$ \log p\sim N(4.15, 0.65^2)$ &$\log q\sim N(7.99,.20^2)$ &$w\sim N(50.5, 43.50^2)$ \\
& $Cor(\log p,\log q)=0.40$ &$ Cor( w,\log q)=0.65$&$ Cor(w, \log p)=0. $\\
&\\
Under $\phi$ :&$\log p\sim N(4.40, 0.65^2)$ & $w\sim N(54.6, 43.50^2)$ & $Cor(w,\log p)=0.$\\
\end{tabular}
\end{table}

\begin{figure}[h]
\centering
\begin{subfigure}[Profit distributions]%{0.5\textwidth}
{\includegraphics[height=0.4\textwidth, width=0.4\textwidth]{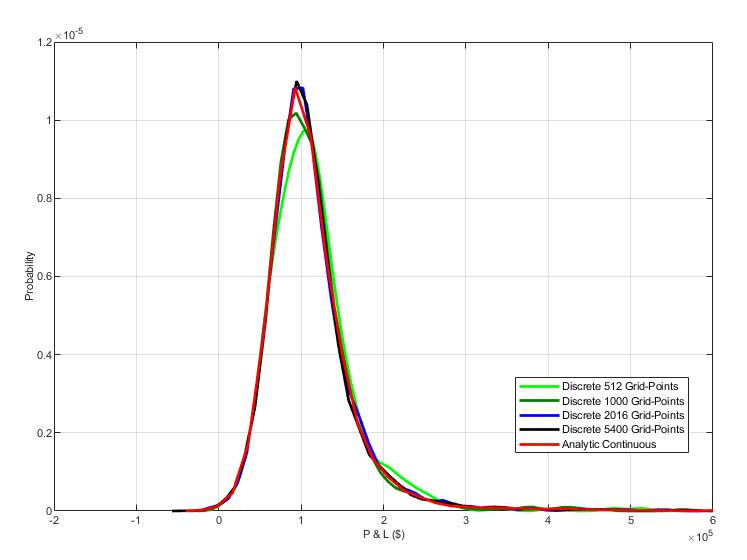}}
\end{subfigure}
\begin{subfigure}[Difference]%{0.5\textwidth}
 {\includegraphics[height=0.4\textwidth, width=0.4\textwidth]{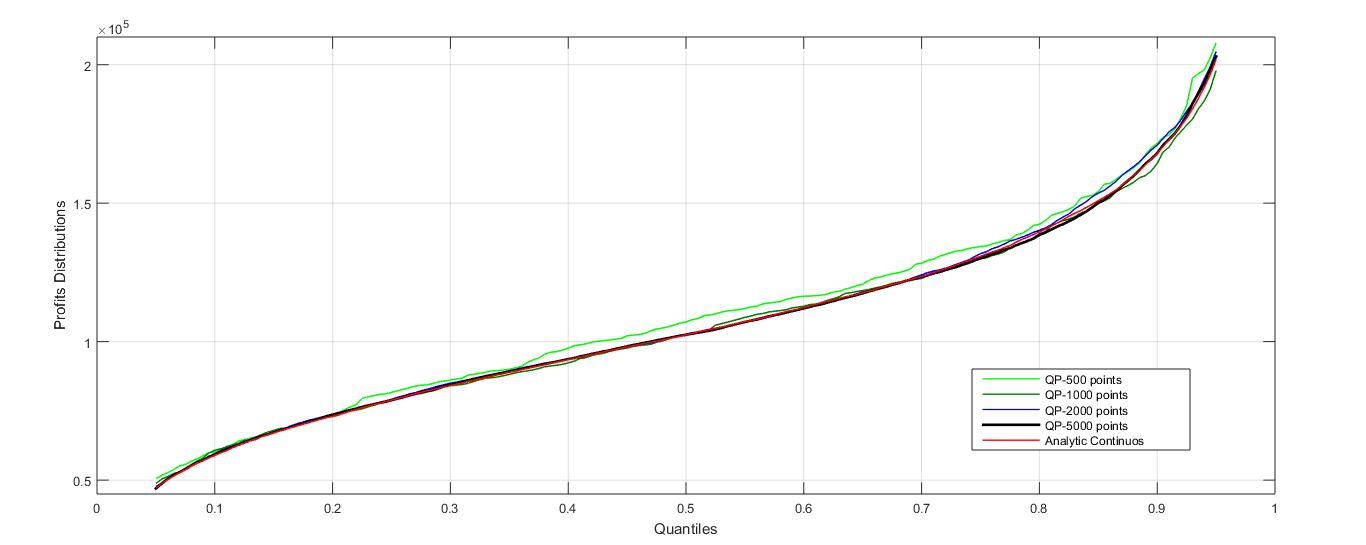}}
\end{subfigure}
\caption{Profit distributions independent case. Continuous vs Discrete solutions.}\label{ffig1}
\end{figure}

Notice that the discrete independence solution can be interpreted as an approximation to the continuous solution. To check the quality of the approximation we vary the number of data points. Besides from $N = 1000$ we take $N= 512$, 2016 and 5400.
As could be observed in Figure \ref{ffig1} the solutions obtained using both methods do not differ much from each other. Figure \ref{ffig1}$(a)$, shows the comparison between continuous and discrete setting using profit distributions with the different number of data-points.
Figure \ref{ffig1}$(b)$ illustrates that difference. We can see that as the number of data-points increases the results converge to the continuous solution. Also for $N=1000$ we already obtain a good approximation, as the discrete and the continuous solutions are very close.

\begin{figure}[h]
\centering
\begin{subfigure}[Profit Distribution]{
\includegraphics[width=0.43\textwidth]{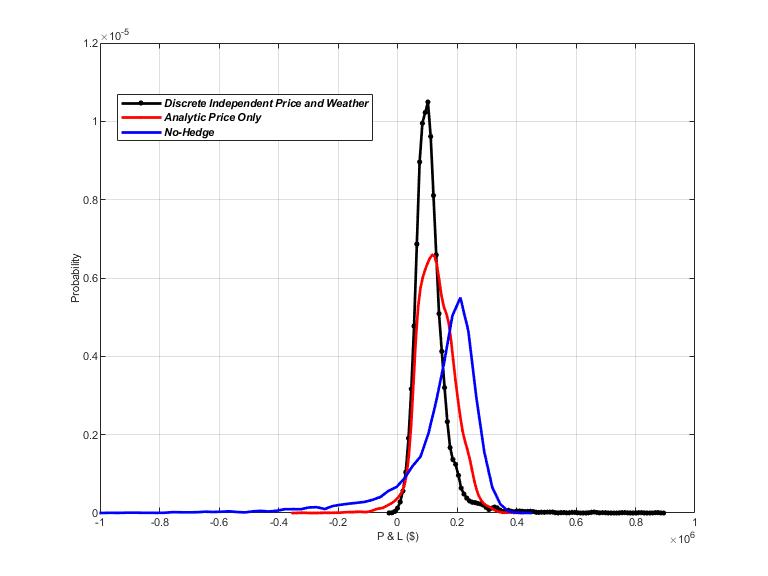}
}
\end{subfigure}
\begin{subfigure}[Profit Quantiles]{
\includegraphics[width=0.41\textwidth]{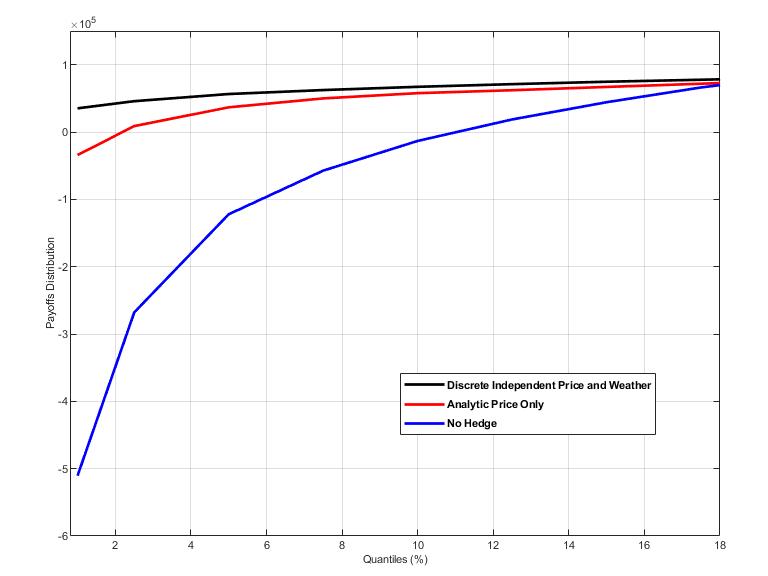}
}
\end{subfigure}
 \caption{Profit distribution in the independence case for different strategies: no-hedge, price only and price plus weather}
 \label{ffig2}
\end{figure}

Next we compare three different strategies for the independence case solution in discrete setting: "No-hedge", "Price Only" and " Price and Weather",  we can see the gain when the weather claim is included into the hedging strategy. Figure \ref{ffig2} shows the profit distributions in the independence case for the three strategies.
We can observe that both hedges reduce the variance of profit relative to the No-hedge strategy.

Table \ref{ttab1} shows the numerical value  of the Quantiles plotted in figure \ref{ffig2}(b). By looking at the table values as well a the figures it can be seen that price plus weather hedge cuts off the left tail of profit distribution compared to the price only hedging. That is the Discrete independence solution allows the ER to protect itself against rare but detrimental events by hedging weather risk.
\begin{table}[htb!]
\caption{ Profit Quantiles in the independence case for different strategies: no-hedge, price only and price plus weather}
%\vspace{-0.5cm}
\begin{center}
\begin{tabular}{|c|c|c|c|}
  \hline
  \multicolumn{4}{|c|}{\textbf{Quantiles}}\\ \hline\hline
   & \textbf{No Hedge} & \textbf{Price Only} & \textbf{Price plus Weather}\\ \hline
  1\% & -508555 & -32799 & 36863 \\ \hline
  2.5\% & -257446 & -1031 & 47979 \\ \hline
  5\% & -115389 & 31788 & 56741 \\ \hline
  7.5\% & -57849& 48827 & 63860  \\ \hline
  10\% & -13519& 56766 & 69667  \\ \hline
  12.5\% & 20731 & 63184 & 73346\\ \hline
  15\% & 39829 & 68587 & 75759 \\ \hline
  17.5\% & 60372 & 73382 & 78544 \\ \hline
  20\% & 86090 & 79253 & 80699\\ \hline
 \end{tabular}\label{ttab1}
\end{center}
\end{table}

\subsection{General case}

In this section we evaluate the proposed methodology in the general case. Without the independence assumption, there is no solution for the continuous case, and thus we only consider the discrete frame.  We use the same data parameters as in the previous section (independence case) except for a positive correlation between $w$ and $p$ (See Table \ref{tab:gen}).

\begin{table}[h]
\caption{Data parameters for the general case.} \medskip
\label{tab:gen}
\begin{tabular}{rlll}
Under $\psi$ : &$ \log p\sim N(4.15, 0.65^2)$ &$\log q\sim N(7.99,.20^2)$ &$w\sim N(50.5, 43.50^2)$ \\
& $Cor(\log p,\log q)=0.40$ &$ Cor( w,\log q)=0.65$&$ Cor(w, \log p)=0.33. $\\
&\\
Under $\phi$ :&$\log p\sim N(4.40, 0.65^2)$ & $w\sim N(54.6, 43.50^2)$ & $Cor(w,\log p)=0.33.$\\
\end{tabular}
\end{table}

\begin{figure}[h!]
\centering
\begin{subfigure}[Profit Distribution]{
\includegraphics[width=0.75\textwidth]{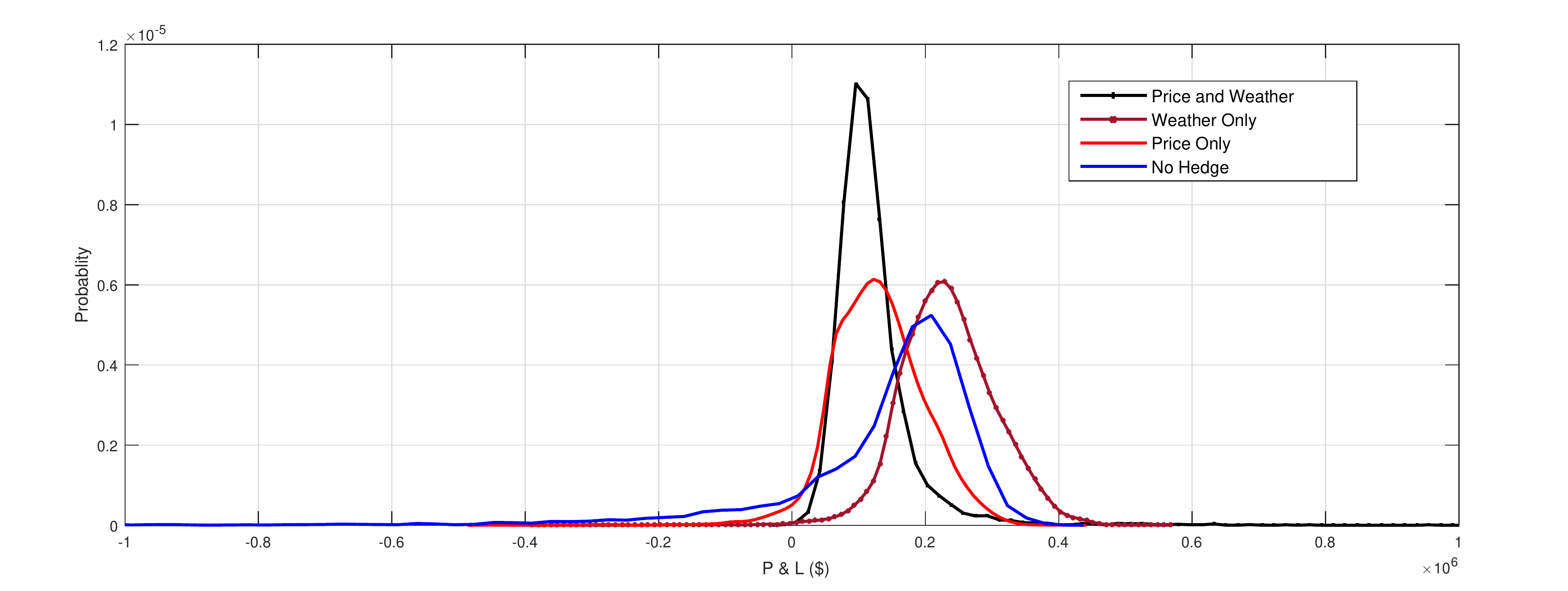}
}
\end{subfigure}
\begin{subfigure}[Profit Quantiles]{
\includegraphics[width=0.75\textwidth]{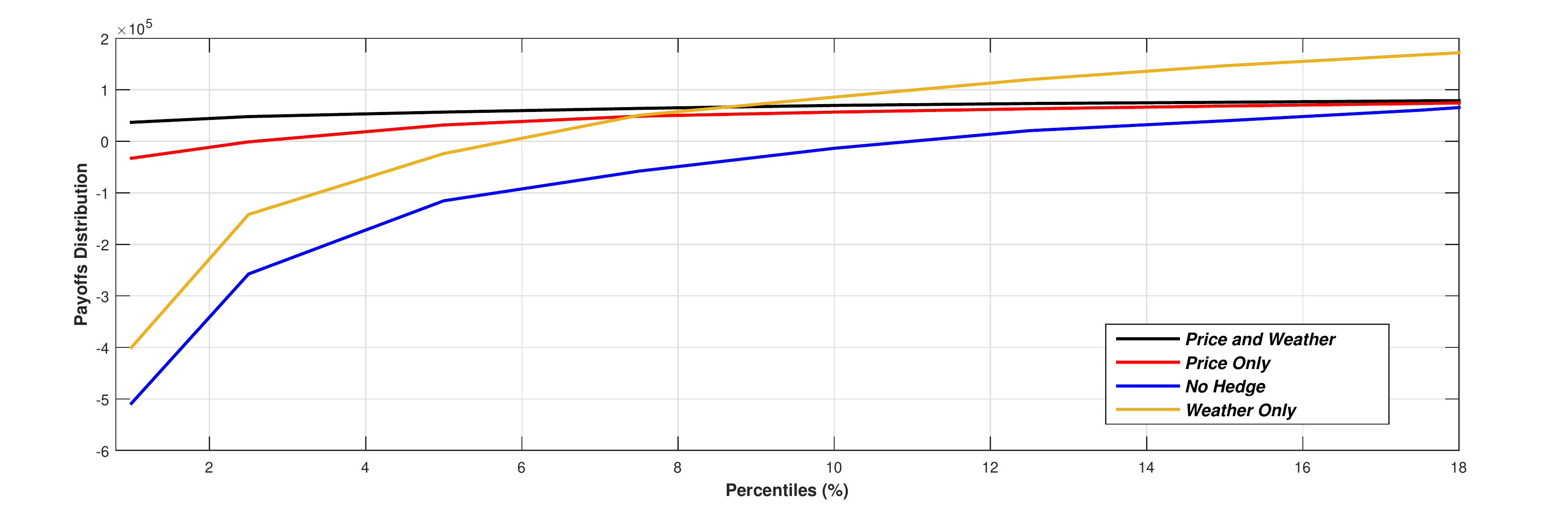}
}
\end{subfigure}
\caption{Profit distributions under the cases: No hedge, Price only, Weather only, and Price and weather.}\label{ffig3}
\end{figure}

Figure \ref{ffig3} illustrates the  comparison of four possible strategies  "No-hedge", "Price Only", "Weather Only" and " Price and Weather".  The weather only strategy seems to dominate in distribution the no-hedge strategy. While the price only has smaller risk, but also has smaller expected profit. As expected, the effect of hedging reduces the variance. When  price and weather instruments are use together  risks  is reduced more than  when only price or weather instruments are used. 

\begin{figure}[h]
\centering
\begin{subfigure}[Price Payoff Tail with different sigma]{
\includegraphics[width=0.8\textwidth]{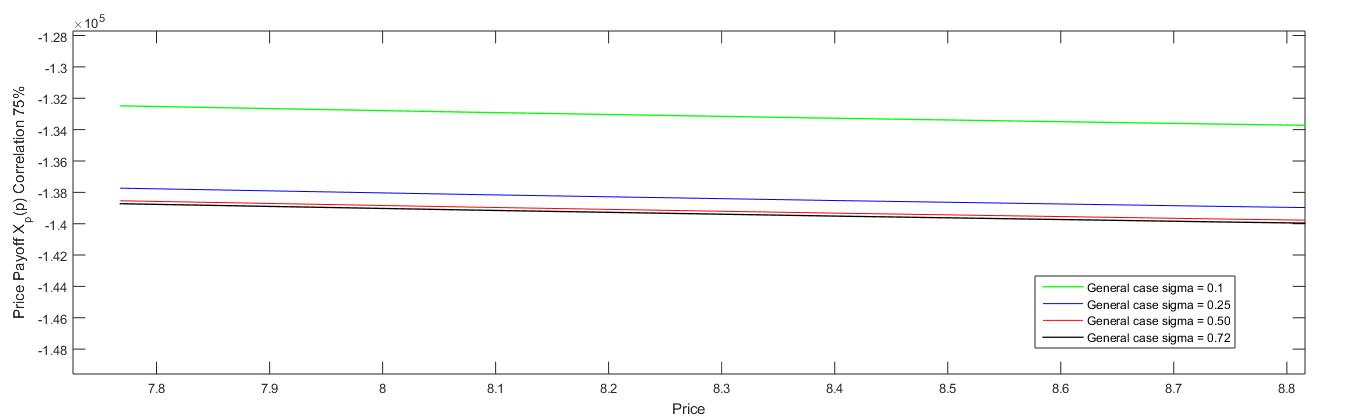}
}
\end{subfigure}
\begin{subfigure}[Weather Payoff Tail with different sigma]{
\includegraphics[width=0.8\textwidth]{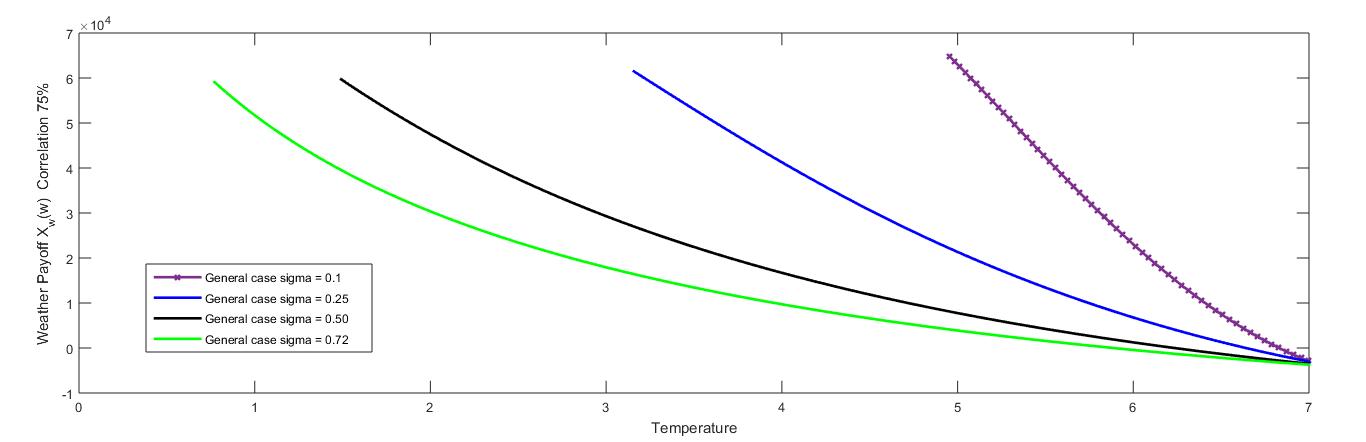}
}
\end{subfigure}
\caption{ Price $X_p(p)$ and $X_w(w)$ Tails for different sigma}\label{ffig6}
\end{figure}

Electricity markets face a higher level of loss risk which conduces to profits that are distributed with heavy tails (see, \cite{AndNom}). The lognormal distributions exhibits fat-tails. In order to analyze how these fat tails affect our models we run a numerical experiment in which we analyze how the tail of both payoffs, price and weather, change due to changes on volatility (sigma = 0.1, 0.25, 0.5 and 0.72) for each price and weather separately, while keeping a fixed correlation value (rho = 0,75).
 In figure~\ref{ffig6} we can observe that the claims (namely the price hedging and weather hedging) show high volatilities levels and if volatility changes that event may produce a heavy tail or extreme value. Still, our model will generate a gain over the other models used as a reference. Figures \ref{ffig6}(a) and (b) show the payoff's fat tails when changing the volatility $\sigma$, for price and weather payoffs respectively.

\subsection{Independence case as proxy to general case}
\cite{Brik_Ron} suggest using the Independence case solution as a proxy to solve the general case. They argue that if the correlation between $p$ and $w$ is small, the solution obtained under the independence assumption is almost optimal for the general case. Next, we analyze the difference between the independence case solution (see Corollary \ref{cor:ind}) and the general case solution (see Theorem \ref{thm:sol}) when the correlation $\rho$ between $p$ and $\log w$ varies from $0\%$ to $75\%$.

First, we analyze how the weather and the price claim change as the correlation between price and weather index vary. As can be seen, while changes in the correlation only produce small changes in the price contingent claim,  the weather contingent claim is affected by the level of correlation between $p$ and $w$. In the weather contingent claim the effect is rather strong with the weather payoff offering a better hedge as correlation increases. Thus when using the independence solution, the price claim will be almost optimal for all values of correlation $\rho$, while the weather claim will lose quality as $\rho$ increases.
\begin{figure}[h]
\centering
\begin{subfigure}[Price Contingent Claim]{
\includegraphics[height=0.3\textwidth, width=0.45\textwidth]{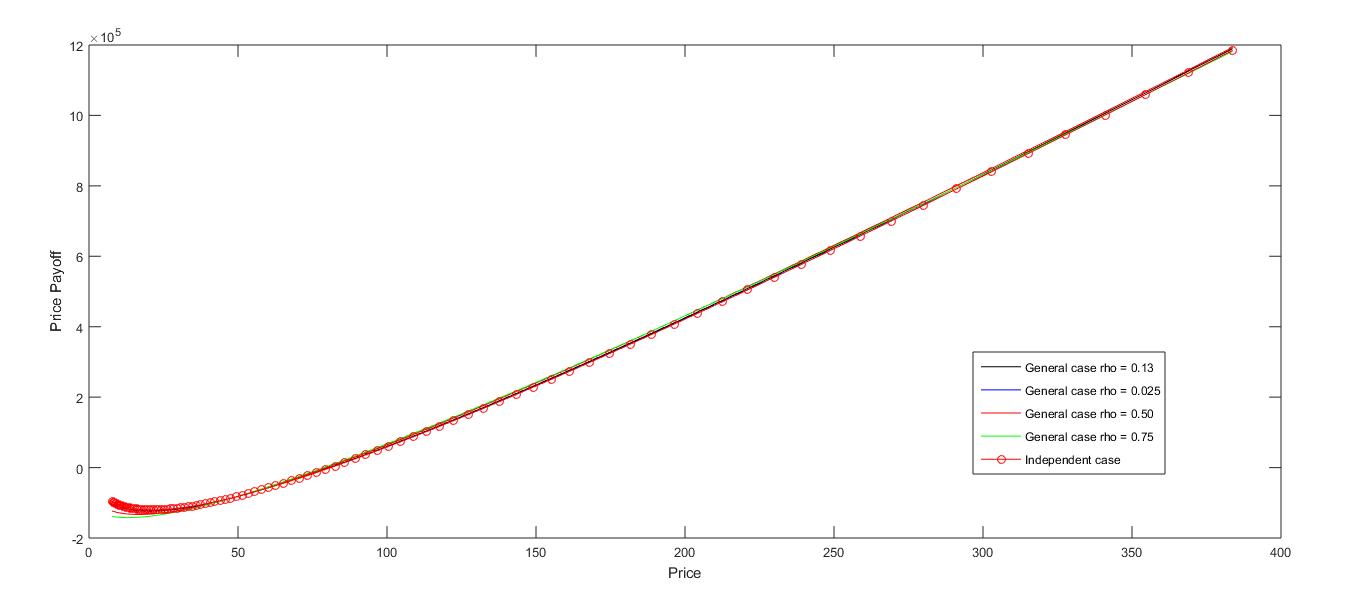}}
\end{subfigure}
\begin{subfigure}[Weather Contingent Claim]{
\includegraphics[height=0.3\textwidth, width=0.45\textwidth] {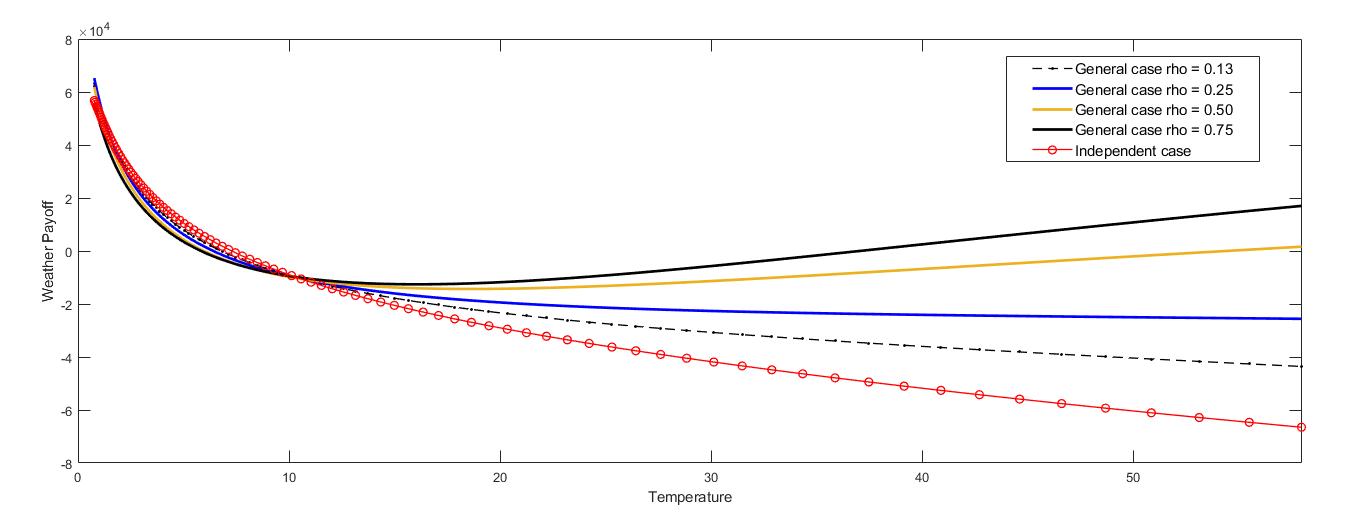}}
\end{subfigure}
\caption{Optimal payoffs for weather and Price, for Independent and General cases with different values of correlation}
\label{ffig4}
\end{figure}

\begin{figure}[h!]
\centering
\includegraphics[width=0.5\textwidth]{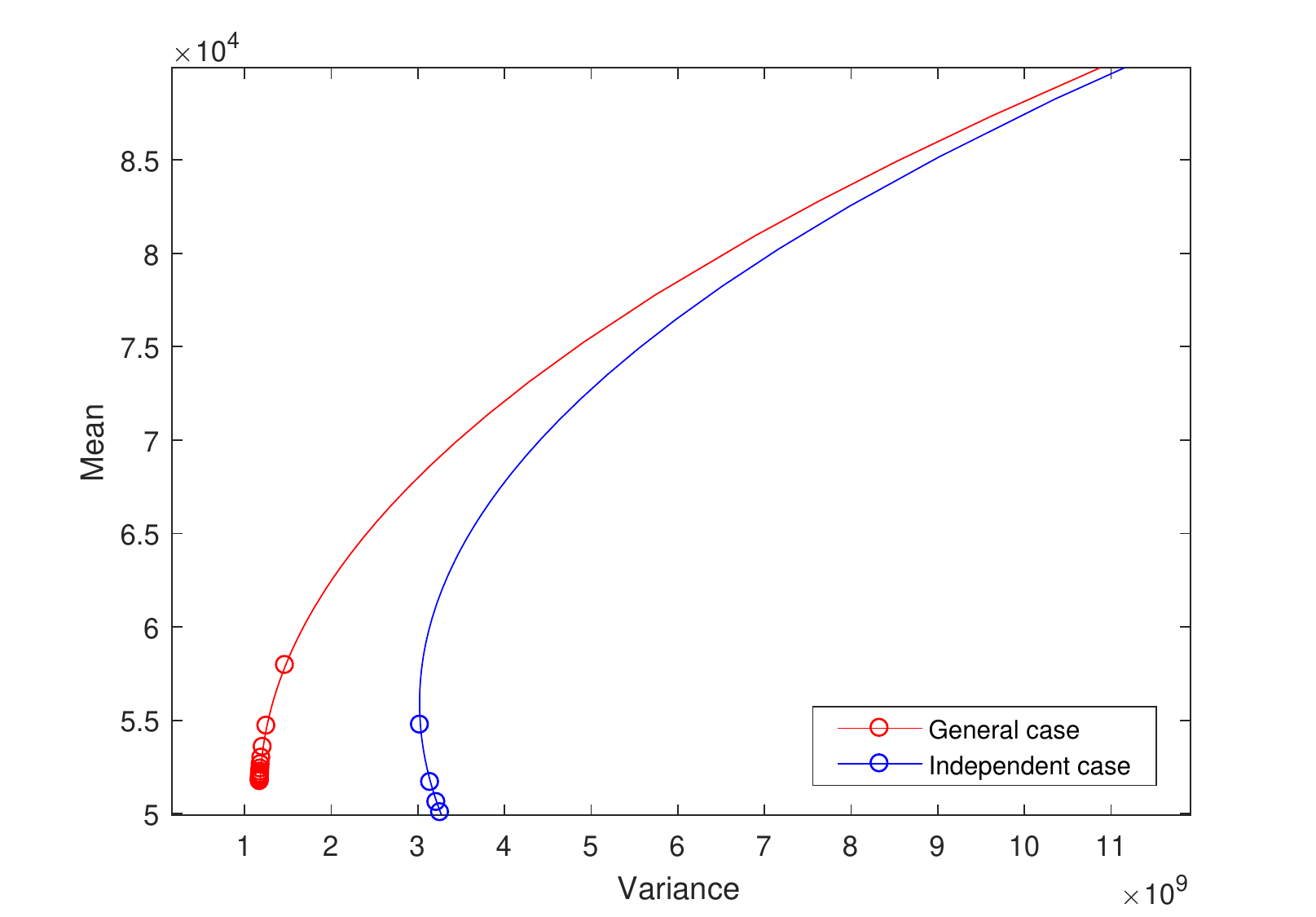}
\caption{ Efficient frontiers for the hedges using price and weather instruments, under the independence assumption and for the General Case}\label{ffig7}
\end{figure}

\begin{figure}[h!]
\centering
\begin{subfigure}[Profit Distribution]{
\includegraphics[width=0.7\textwidth]{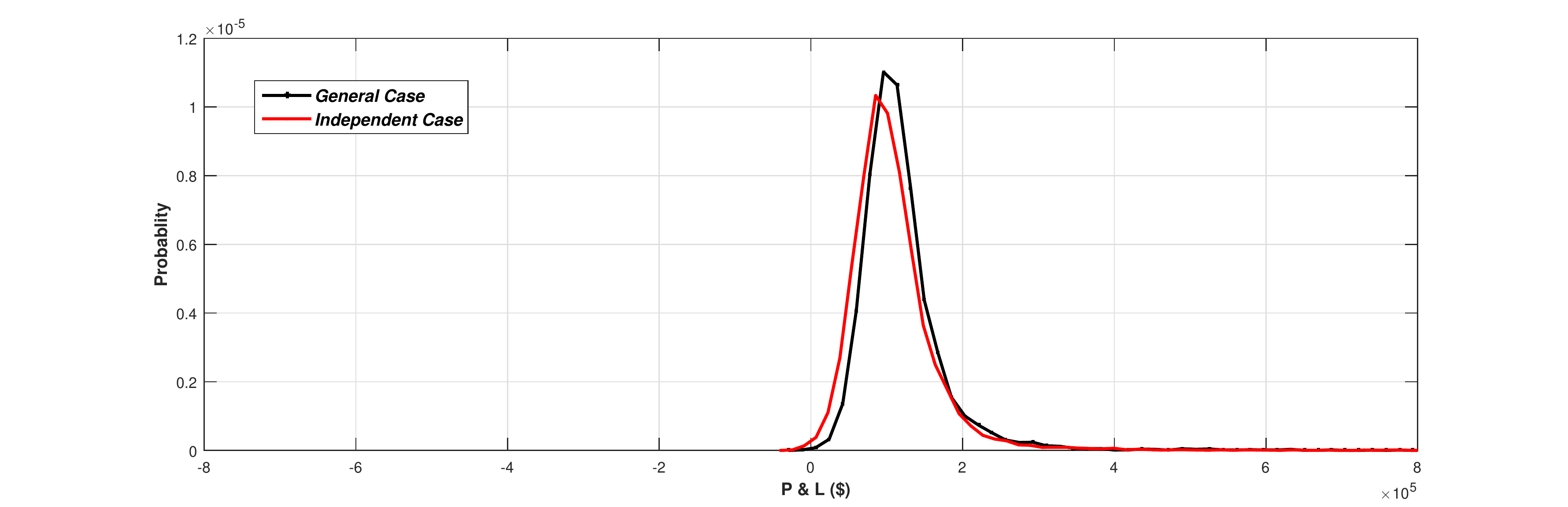}
}
\end{subfigure}
\begin{subfigure}[Profit Quantiles]{
\includegraphics[width=0.7\textwidth]{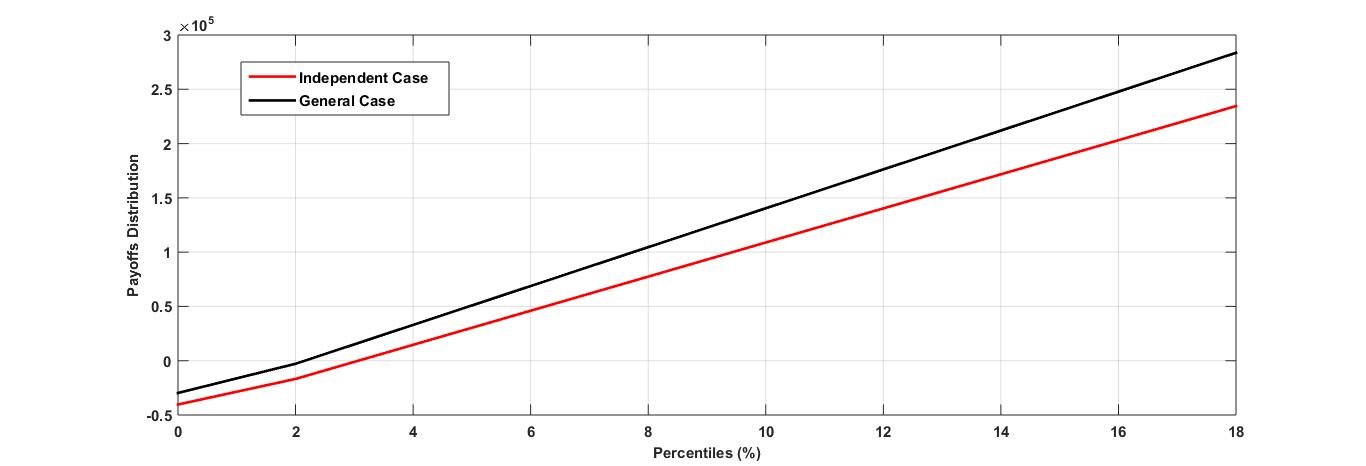}
}
\end{subfigure}
\caption{ Profit distribution and percentiles for the hedges using price and weather instruments, under the independence assumption and for the General Case}\label{fig:GenInd}
\end{figure}

We can see that the general solution completely dominates the one constructed under the independence assumption. This fact is illustrated, in Figure \ref{ffig7}, where the efficient frontier for the optimal solution of the general case (see Corollary \ref{prop:frontier}) and for the independence case solution are presented. To see the effects on the hedge,
Figure \ref{fig:GenInd} shows the gain of the General case over the Independence one in terms of profit distribution and Quantiles.

\begin{figure}[h!]
\centering
\includegraphics[width=0.75\textwidth]{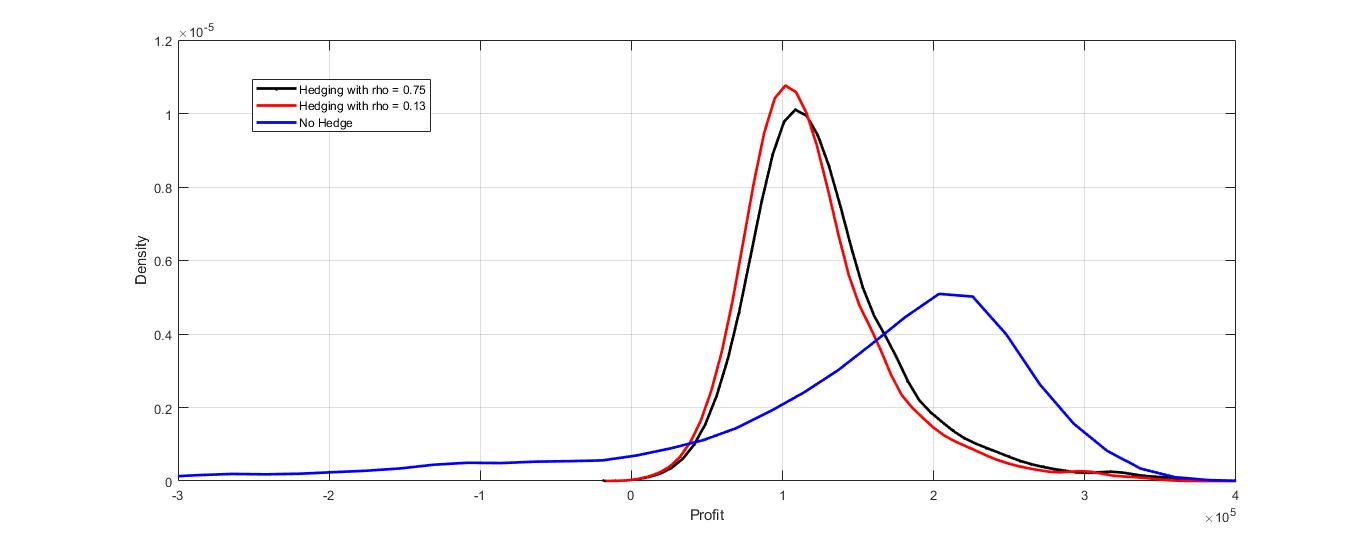}
\caption{ Hedged profit for different level of correlation between $p$ and $w$ in the General Case solution}\label{ffig10}
\end{figure}

In Figure \ref{ffig10} hedge profit distributions under different levels of correlation $\rho = 0.13$ and $\rho=0.75$ are compared.
The figure illustrate that similarly to what is observed in the comparison between the general and independence case, the hedge improves as the level of correlation between price and weather increases.

\section{Conclusions}\label{sec5-2}

This paper develops numerical methods to determine the optimal derivative contingent claim written on both electricity price and a weather; aiming to improve the performance of the hedging claim due to the link between price, demanded quantity and weather-linked index.  We introduce a discrete framework which allows to construct optimal contingent claims as functions of price and weather, without any assumption on the underlying distributions. Our solution method is based on Quadratic Optimization; we give a closed form solution for the basic model consider here, where only zero cost constraints  on the contingent claims are considered. But, the Quadratic Optimization model could be applied under more general linear and second order constraints, in which case numerical solutions could be obtained.

Our method improves upon the existing literature, which has only considered the case when weather linked index and price are independent (\cite{Brik_Ron} and \cite{Pant11}).  Our numerical results illustrate the gain due to the inclusion of the weather variable, improving on existing hedging positions. Our results confirm that the weather contingent claim allows adjustment of hedge strategy with the price contingent claim in order to hedge the double exposure of the agents. Our numerical results support the gain of the proposed strategy over the best performing claim derived for strategies using price instruments only and over strategies using price and weather instrument assuming independence of price and weather.

We derive several results characterizing the optimal contingent claims. In particular we show the existing of a `two fund' theorem in this case. Also, we show that when the market measure and the real world measures coincide the optimal solution to the given model, for any $a \ge 0$  is the contingent claim that minimizes the total profit risk and maximizes the expected total profit.

\pagebreak
\bibliographystyle{chicago}

\end{document}